\documentclass[%
 aip,
 amsmath,amssymb,
 reprint,%
]{revtex4-2}
\usepackage{graphicx}
\usepackage{dcolumn}
\usepackage{bm}

\usepackage[utf8]{inputenc}
\usepackage[T1]{fontenc}
\usepackage{mathptmx}
\usepackage{etoolbox}

\makeatletter
\def\@email#1#2{%
 \endgroup
 \patchcmd{\titleblock@produce}
  {\frontmatter@RRAPformat}
  {\frontmatter@RRAPformat{\produce@RRAP{*#1\href{mailto:#2}{#2}}}\frontmatter@RRAPformat}
  {}{}
}%
\makeatother
\begin{document}
                                                                                                                                                                                                                                                                                                                                                                                                                                                                                                                                                                                                                                                                                                                                                                                                                                                                                                                                                                                                                                                                                                                                                                                                                                                                                                                                                                                                                                                                                                                                                                                                                                                                                                                                                                                                                                                                                                                                                     
\title{Hot-carrier thermal breakdown and S-type current-voltage characteristics in
 perforated  graphene  structures
}
\author{V.~Ryzhii$^{1,2*}$,  C.~Tang$^{1,2}$,   
M.~Ryzhii$^{3}$, and M. S. Shur$^4$
}
\address{$^1$Frontier Research Institute for Interdisciplinary Sciences,
Tohoku University, Sendai 980-8578, Japan\\
$^2$Research Institute of Electrical Communication,~Tohoku University,~Sendai~ 980-8577,
Japan\\
$^3$School of Computer Science and Engineering, University of Aizu, Aizu-Wakamatsu 965-8580, Japan\\
$^4$Department of Electrical, Computer, and Systems Engineering,\\ Rensselaer Polytechnic Institute,~Troy,~New York~12180,\\ USA\\
*{Author to whom correspondence should be addressed: vryzhii@gmail.com}
}
\begin{abstract}
We investigate the carrier transport characteristics of perforated graphene layer (PGL) composed of arrays of interdigital coplanar graphene microribbons (GMRs) connected by graphene nanoribbon (GNR) bridges. We analyze their operation at room-temperature. Under an applied bias voltage, two-dimensional electron and hole systems (2DES and 2DHS) form in adjacent GMRs. The terminal current in these PGL structures is primarily governed by thermionic transport across the GNR bridges. As electrons and holes traverse the GNRs, they induce heating in the 2DES and 2DHS, creating a positive feedback loop between carrier heating and thermionic emission. This phenomenon, characterized as hot-carrier thermal breakdown, can give rise to S-shaped  inter-GMR current-voltage characteristics. These unique transport properties make PGLs promising candidates for fast, voltage-controlled room-temperature switches and electromagnetic radiation detectors.
\end{abstract}


\maketitle


Remarkable properties of graphene layers (GLs),  enable their use indetectors, amplifiers, and sources of the terahertz radiation,~\cite{1,2,3,4,5,6,7,8,9,10,11,12,13,14,15,16,17}  as well as optoelectronic devices,~\cite{18,19,20} and other systems based on graphene micro- and nanoribbons (GMRs and GNRs)
and hybrid structures.~\cite{21,22,23,24,25,26,27,28} 
The graphene-based topological structures using GL-GMR-GNR arrays presents
 new opportunities  for enhancing of the performance of graphene devices.
 Recently, GMR-GNR terahertz  photomixers and detectors based on in-plane interdigital GMR-GNR arrays, where GNR bridges connect adjacent GMRs to form inter-GMR barriers through bandgap opening,\cite{29,30,31} have been proposed and evaluated.~\cite{32,33} 
 Such or similar device structures can be fabricated 
on the perforated graphene layers (PGLs)  
with the GNRs
bridges connecting the GNRs.
The energy barriers in the inter-perforation constrictions
depend on the perforation shape  and carrier transport across the GNR bridges can occur via tunneling, thermo-assisted tunneling, or thermionic mechanisms.
The  dominant transport mechanism is critical for the operation of different devices with the GNR bridges.
In constrictions with  variable width characterized by fairly smooth barriers (such as parabolic barriers), the energy barriers are relatively smooth. In such cases, the thermionic mechanism can play a significant role even at large band openings ($2\Delta$), provided that the terminal voltage $V_G$  is sufficiently high. 

\begin{figure}[t]
\centering
\includegraphics[width=7.5cm]{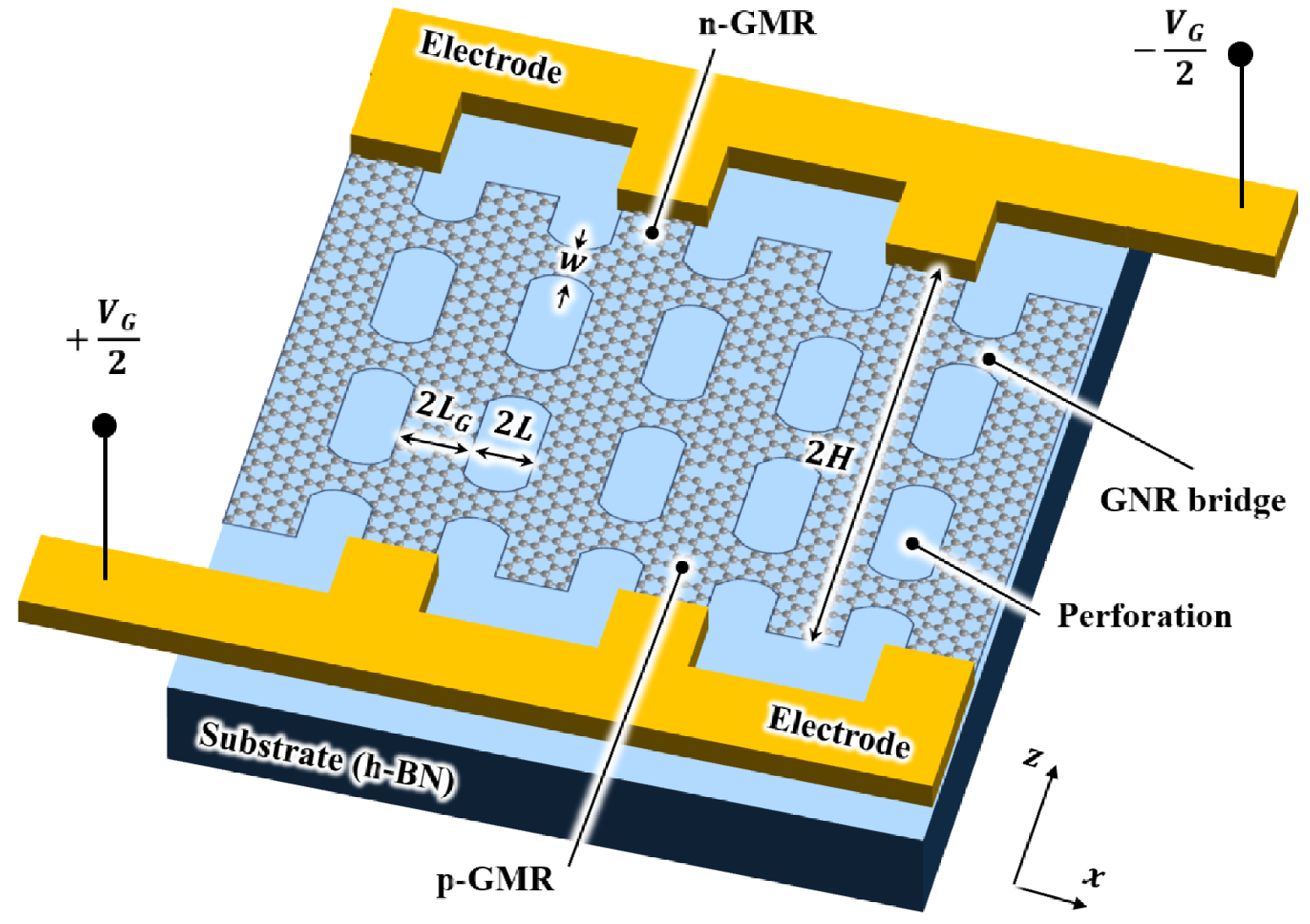}
\caption{Schematic top view of the PGL device structure. 
}\label{Fig1}
\end{figure}
In this paper, we consider  the transport  in the PGLs featuring 
interdigital coplanar GMRs connected by the GNR bridges of nearly parabolic form.
These PGLs can exhibit interesting features the occurrence of the hot-carrier thermal breakdown leading to the S-shaped dependences of the electron and hole  effective temperature $T$ and the terminal current $J$ on the bias voltage $V_G$ (the $T-V_G$ and $J-V_G$ characteristics).
Notably, the proposed  PGLs  can serve  as  fast voltage-pulse-controlled
current
 switches  operating at room temperature.

\begin{figure}[t]
\centering
\includegraphics[width=7.5cm]{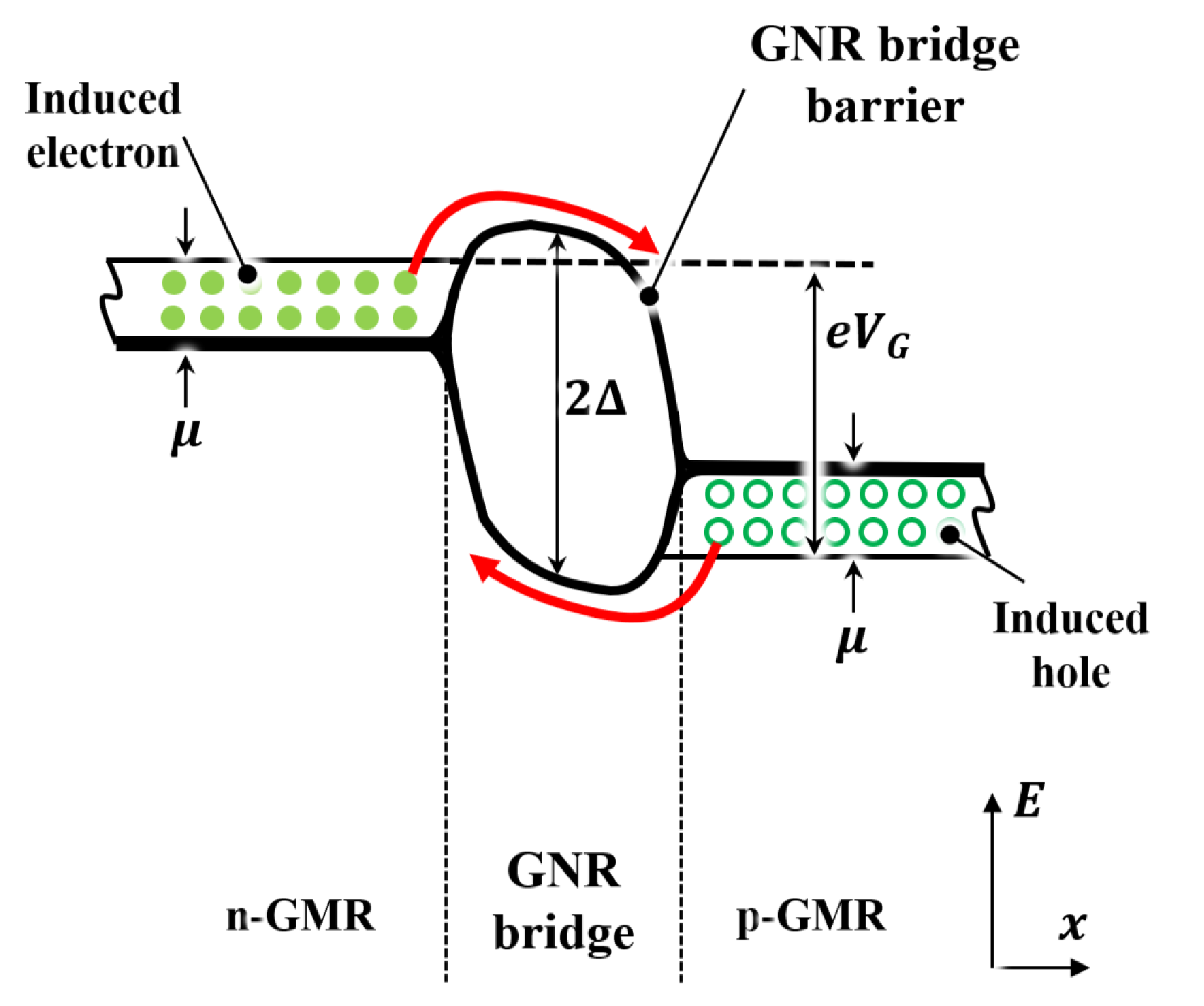}
\caption{The PGL   band diagram  for one structure  period at  cross-sections corresponding to GNRs.
}\label{Fig2}
\end{figure}


Figure~1 shows the PGL structure  under consideration.
This structure is  an array of the  interdigital strips (i.e., GMRs)
connected  by GNRs bridges.
The opposite ends  of the interdigital GMRs  are connected to the corresponding poles of the bias voltage source. 
The terminal voltage $V_G$ (applied to the ends
of the neighboring GMRs) induces the electron and hole charges, i.e., the two-dimensional electron and hole systems (2DESs and 2DHSs),
 in the  neighboring GMRs.  
Therefore  the n-GMR and p-GMRs alternate. The PGL structures are assumed to be placed on  substrates, such as  the hexagonal-BN substrates or embedded in the layers of this material. 
Figures 2  presents the band diagram at the PGL structure at the cross-sections corresponding
 to the GNRs under the applied voltage $V_G$.  
We assume that the minimal thickness  of the GNRs  is chosen  to be  sufficiently small to provide a reasonable band gap opening due to quantization of the electron and hole spectra.
The number, $2N-1$, of the CNR bridges between each pair of the GMRs may vary ,
provided it satisfies  the condition $(2N-1W) \ll 2H$, where $2H$ is the GMR length.
The GNR width $2L$ should be  sufficiently large to  prevent the quantum coupling
of the electrons and holes in adjacent GNRs.
These constrains  imply that the inter-GMR
conductance (through all the GNRs) remains much smaller than  the conductance along the GMRs.
This assumes that the energy barriers for the electrons and holes in the perforations between the GNRs
are sufficiently large,  being small compared with the energy barriers in the perforations
(determined by the GMR/h-BN band offsets).
As a result, the electrons and holes  incident on these barriers are reflected (at least atmoderate voltages),
 so that the inter-GMR currents through the perforations are  suppressed. 
Consequently, the  electron and hole currents between the neighboring GMRs
flow through the GNR bridges.

The GMR number, $2M$, can vary from two ($M =1$, i.e., one n-GMR and one p-GMR) to a  large number 
($M \gg 1$). 
Apart from the applied voltage $V_G$, the electrically induced electron and hole densities, depend on
the inter-GMR capacitance $c_G$,~\cite{34,35}
 which, in turn, is determined by  the GMR width  $2L_G$ and the inter-GMR spacing equal to $2L$ (see Fig.~1).
Due to the in-plane configuration, the inter-GMR capacitance $c_G$ can be much smaller than the capacitance
of the gated GMRs (as in the field-effect transistors). The capacitance smallness is beneficial for enhancing the speed of the  PGL-based devices.

The height of the barriers in the GNR bridges   (limiting the electron and hole current through them),
formed due to the electron and hole transverse confinement in the GNR  and, hence,  the quantization of the electron and hole spectra 
is estimated as $\Delta = \pi\hbar\,v_W/w$. Here 
$w$ is the GNR minimal width ($w < W$ or even $w \ll W$), 
 $v_W \simeq 10^8$~cm/s is the carrier velocity in GLs, and $\hbar$ is the Planck constant.
 We assume that the  GNR width $W(x)$ (with $W(x)|_{x=0} = w$) corresponds to the
 parabolic form  of the energy barrier $\Delta(x)$, such that  $\Delta(x)|_{x=\pm L} =0$ and $\Delta(x)|_{x=0} = \Delta$.
 
\begin{table}[b]
\centering 
\vspace{2 mm}
\begin{tabular}{|r|c|}
\hline
\hline
GMR length,\,  $2H$& (0.5 - 1.0)~$\mu$m\\
\hline
GNR length,\,  $2L$ &  40~nm\\
\hline
 GMR width, $2L_G$ &  60~nm\\
\hline
GNR minimal width,\, $w$ &  (6 - 12)~nm\\
\hline
GNR barrier height, $\Delta$ &  (150 - 300)~meV\\
\hline
Substrate dielectric constant,\, $\kappa_S$ &  10\\
\hline
GMR-GNR capacitance, $c_G$& 0.576\\
\hline
Characteristic voltage, ${\overline V}_G$ & 8 meV\\
\hline  
Number of GNR bridges, $(2N-1)$ & 1 and 5 \\ 
\hline
Lattice temperature,\, $T_0$&\, 25~meV \\ 
\hline
Optical phonons energy,  $\hbar\omega_0$& 200~meV \\
\hline
OP energy relaxation   time, $\tau_0^{\varepsilon}$,& 20~ps\\
\hline
SC energy relaxation parameter, $S$ & 10\\
\hline
Carrier momentum relaxation frequency,\,  $\nu$&\, 1 and 3~ps$^{-1}$\\ 
\hline
\hline
\end{tabular}
\caption{\label{table} Main device parameters} 
\end{table}

\begin{figure*}[t]
\centering
\includegraphics[width=12.0cm]{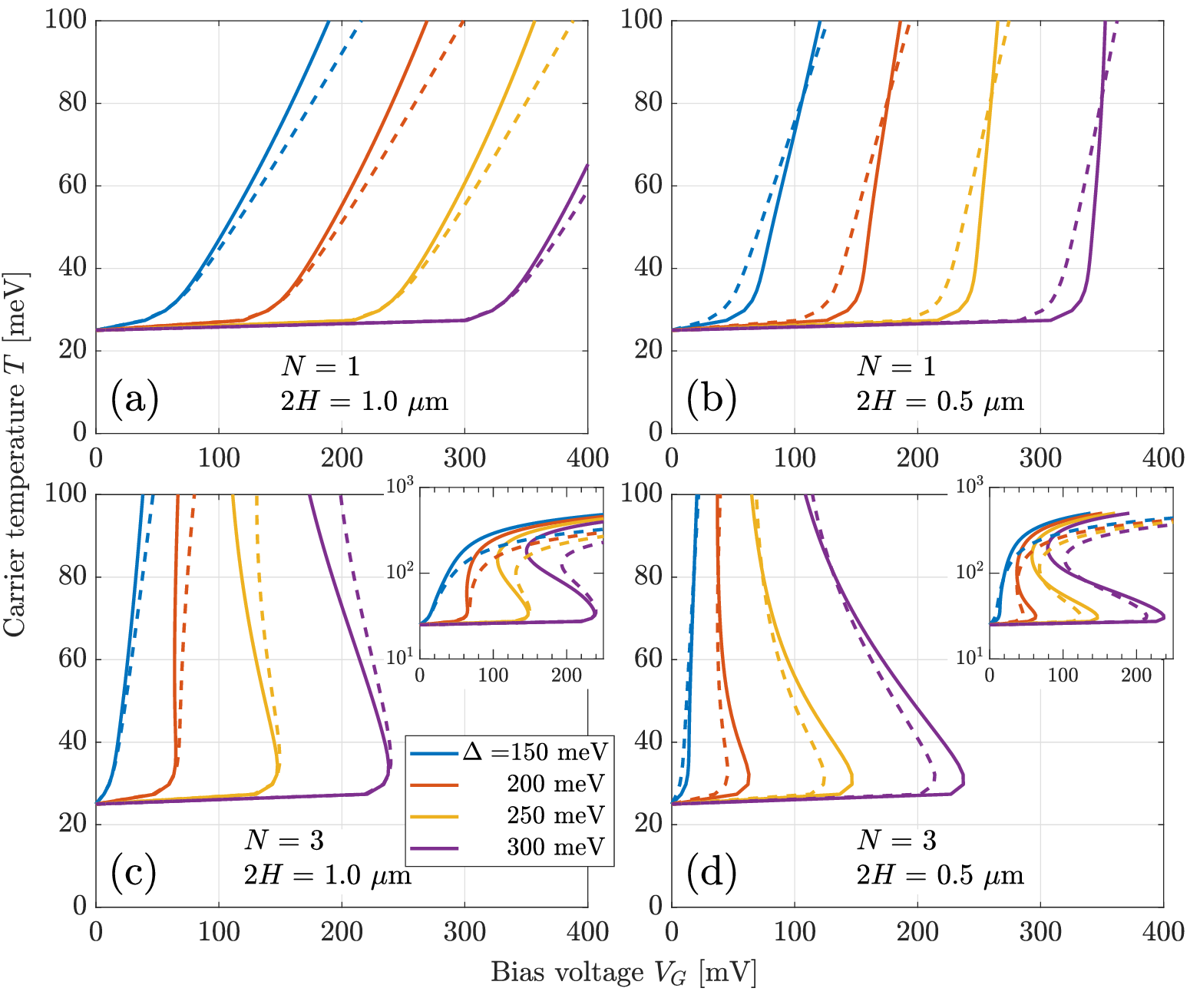}
\caption{The $T-V_G$ dependences
for devices with different structural parameters
 for  the carrier collision frequency $\nu = 1$~ps$^{-1}$ - solid lines   and $\nu = 3$~ps$^{-1}$ - dashed lines, respectively.
 The insets show the same dependences as in (c) and (d) in a wider  temperature range   (in the logarithmic scale).}
\label{Fig3}
\end{figure*}

\begin{figure*}[t]
\centering
\includegraphics[width=12.0cm]{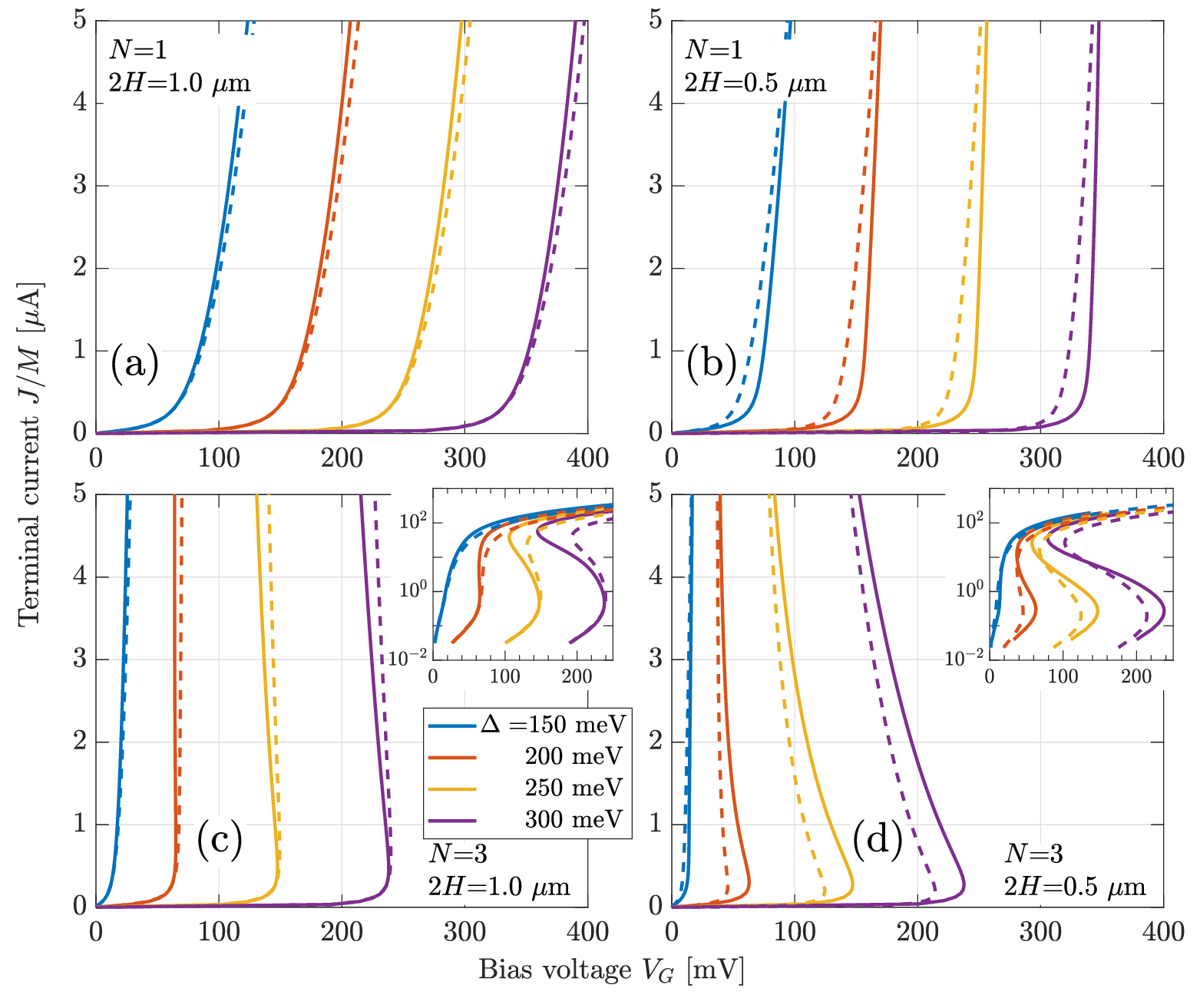}
\caption{The terminal current $J$  (normalized by the number, $M$, of GMR pairs) versus bias voltage $V_G$ for the $T-V_G$ dependences  in Fig.~3:  $\nu = 1$~ps$^{-1}$ - solid lines   and $\nu = 3$~ps$^{-1}$ - dashed lines. {REPLACE H by 2H mum}
}
\label{Fig4} 
\end{figure*}

Following the Landauer-Buttiker formula~\cite{36} (see also, for example, Refs.~\cite{37,38,39}) applied to the one-dimensional transport through the GNRs,
the current, $j_{GNR} = j_{GNR}(z)$, of the electrons and holes via one  GNR, forming a near-parabolic energy barrier, at the net voltage $V_G$ between its ends, can be approximated by the following extrapolation:

\begin{eqnarray}\label{eq1}
j_{GNR} \simeq \frac{8eT}{\pi\,\hbar}\exp\biggl(\frac{\mu-\Delta}{T}\biggr)\sinh\biggl(\frac{eV_G}{2T}\biggr).
\end{eqnarray}
Here 
 $\mu = e\sqrt{{\overline V}_GV_G}$ is the electron and hole Fermi energy in the pertinent GMRs induced the bias voltage with the characteristic voltage ${\overline V}_G = (\pi\,c_G\hbar^2v_W^2/2e^3L_G)$, where
  $c_G= [(\kappa_S+1)/4\pi^2]{\overline c}_G$ is the inter-GMR capacitance, $\kappa_S$ is the substrate dielectric constant, and ${\overline c}_G$ is a function of the $L_G/L$ ratio . 


The injected carriers transfer the energy $\delta \varepsilon = eV_G$ per  carrier to the respective  GMR 
Considering this,
the carrier effective temperature $T$,
 can be determined by balancing  the energy received by the 2DES and 2DHS in the GMRs
due to the injection of the hot carriers, and the energy 
dissipated 
to the lattice and the side contacts. 
Hence, the carrier energy balance in the GMRs is governed by
the following equation:

\begin{eqnarray}\label{eq2}
4MHL_G\Sigma_G R
= JV_G.
\end{eqnarray}
Here
 $\Sigma_G = c_GV_G/e$
is the GMR carrier density induced by the bias voltage and $J =M(2N-1j_{GNR}$ is the net terminal current.
The quantity $\Sigma_G R$ is the rate of  carrier energy relaxation in the GMRs per unit of their area. 
We focus on  the PGL at room temperature, when  
the carrier energy relaxation on optical phonons is a crucial mechanism (see, Refs.~39-55) . However, at the high carrier temperatures comparable to or higher than the optical phonon energy in GMRs $\hbar\omega_0 \simeq 0.2$~eV, the rate of  energy   relaxation due to optical phonon mechanism (OP mechanism)) saturates  becoming unable to compensate
the carrier heating associated with the hot carrier injection (see below). This implies that at elevated carrier temperature other energy relaxation mechanism can be essential. Due to this, apart from 
the optical phonon scattering mechanism (the OP mechanism), 
 we include in our model the energy relaxation mechanisms associated with   the supercollisions and the plasmonic interactions~\cite{56,57,58} (the SC mechanism) and the mechanism associated with .
 the transmission of the carrier heat to the side contact and their cooling at the contact (CC mechanism),~\cite{59,60,61,62}. As a result,
the energy balance is described by the following equation (compare with Refs.~10 and 54):

\begin{eqnarray}\label{eq3}
 R \simeq \frac{\hbar\omega_0}{\tau_0^{\varepsilon}}\biggl(\frac{T_0}{\hbar\omega_0}\biggr)^2
\biggl[
\exp\biggl(\frac{\hbar\omega_0}{T_0}-\frac{\hbar\omega_0}{T}\biggr) - 1\biggr]
\nonumber\\
 + \frac{1}{\tau_S^{\varepsilon}}\frac{(T^3-T_0^3)}{3T_0^2} + \frac{(T-T_0)}{\tau_C^{\varepsilon}}.                                                                                                                                                                                                                                                                                                                                                                                                      
\end{eqnarray}
Here 
 $ \tau_0^{\varepsilon}= \tau_0(T_0/\hbar\omega_0)^2 \exp(\hbar\omega_0/T_0)$, 
 $\tau_S^{\varepsilon}= S/\nu$, and  $\tau_C^{\varepsilon} = 2\nu(2H)/v_W^2$ 
 are the "warm" carrier energy relaxation times (i.e., at $ T \simeq T_0$) 
due to the OP, SC, and CC mechanisms, respectively, $\tau_0$ is the characteristic time of the spontaneous optical phonon emission,  $\nu $ is the carrier collision frequency on acoustic phonons, impurities, and due to carrier viscosity effects~\cite{63}, with $S > 1$ being considered as the fitting parameter). 
At $ T \simeq T_0$, Eq.~(3) yields  $ R \simeq (T-T_0)/\tau^{\varepsilon}$ with $1/ \tau^{\varepsilon} = 1/\tau_0^{\varepsilon}+1/\tau_S^{\varepsilon} +  1/\tau_C^{\varepsilon}$.

 Equations~(1), (3), and (4) lead to
  the following equations parametrically relating $J$ and $V_G$ via $T$ as a parameter:

\begin{eqnarray}\label{eq4}
J \simeq J_M
\biggl[\exp\bigg(\frac{\hbar\omega_0}{T_0}- \frac{\hbar\omega_0}{ T}\biggr)-1 
+ 
\frac{(T^3-T_0^3)}{T_0^3}K_S+ \frac{(T-T_0)}{T_0}K_C
\biggr]
\end{eqnarray}
and
\begin{eqnarray}\label{eq5}
V_G = \frac{{\overline V}_G}{4}\biggl[\sqrt{1+\frac{4B(T)}{e{\overline V}_G}}-1\biggr]
^2.
\end{eqnarray}
The solution of   coupled Eqs.~(5) and (6) yields the $T-V_G$ and $J-V_G$ characteristics.
The function $B(T)$ in Eq.~(6) is given by

\begin{eqnarray}\label{eq6}
B(T) = 2T \sinh^{-1}\biggl\{\frac{T_0}{2T\Theta_N} \exp\biggl(\frac{\Delta}{T}\biggr)\nonumber\\ 
\times\biggl[\exp\bigg(\frac{\hbar\omega_0}{T_0}- \frac{\hbar\omega_0}{ T}\biggr)-1
 +  
\frac{(T^3-T_0^3)}{T_0^3}K_S+ \frac{T-T_0}{T_0}K_C
\biggr]\biggr\}.
\end{eqnarray}
Above
$J_M =(4MHc_GT_0)/e\tau_0^{\varepsilon})(T_0/\hbar\omega_0)$ is the characteristic current, 
 $K_S = (\tau_0^{\varepsilon}/3\tau_S^{\varepsilon})(\hbar\omega_0/T_0)$ and $K_C =(\tau_0^{\varepsilon}/\tau_C^{\varepsilon})(\hbar\omega_0/T_0)$ describe the relative contribution of the energy relaxation mechanisms under consideration,
and  $\Theta_N = (2N-1)
 [{2\tau_0^{\varepsilon}e^2/\pi\hbar\,Hc_G}](\hbar\omega_0/T_0)$.


At not too large voltages when $\sqrt{{\overline V}_G}, V_G < T_0/e$
the carrier effective temperature
only slightly exceeds the lattice temperature, i.e., $ T \gtrsim T_0$. In this case, Eqs.~(4) - ( 6) yield     
 ~\cite{11}

\begin{eqnarray}\label{eq7}
 J \simeq 
4M(2N-1)\biggl(\frac{e^2}{\pi\hbar}\biggr)\exp\biggl(-\frac{\Delta}{T_0}\biggr)\,V_G
.
\end{eqnarray}

Further increase in the terminal voltage leads to a steep rise of the carrier temperature
and the terminal current. 
It is interesting that
when   parameter $\Theta_N$  is sufficiently large (there are several GNR bridges between the GMRs and the GMR length $2H$ is not too long),
the derivative $dT/dV_G$ 
 turns to infinity at certain threshold voltage ${\overline V}_{N}^{th}$, which corresponds to the threshold temperature ${\overline T}N^{th}$. 
According to Eq.~(4), the differential conductance
$d J/dV_G \propto 
dT/d V_G$,
 $dJ/dV_G$  also turns to infinity at the same threshold voltage ${\overline V}_{N}^{th}$. As follows from Eqs.~(5) and (6), the threshold voltage ${\overline V}_{N}^{th}$ exists if the parameter $\Theta_N \propto (2N-1)/H$ is sufficiently large, exceeding a certain critical value $\Theta$.

Beyond the voltage threshold, an increase in the effective temperature and the current can not be limited by the OP energy relaxation mechanism.  This phenomenon can be referred to as the hot-carrier thermal breakdown.
In this case,
 the  $T-V_G$ and $J-V_G$ characteristics can be ambiguous

In such a situation, the rise of $T$ and $J$ at $V_G > {\overline V}_{N}^{th} $
is stopped by the inclusion of the SC and CC  mechanisms, which are  effective at elevated carrier temperatures,  so that
 the  $T-V_G$ and $J-V_G$ characteristics shape becomes S-type.

Figures~3 and 4 show the $T-V_G$ and $J-V_G$ characteristics  obtained using the above equation for the PGL structures with different numbers of the GNR bridges $(2N-1)$, different GMR length $2H$,    different values of the GNR energy barrier $\Delta$, and 
collision frequency $\nu$.
 Table I lists the device parameters used in the calculations.

 One can see from Figs.~3(a) and 3(b) that in the structures with a single GNR bridge [$(2N-1) = 1$ and $H = 1.0~\mu$m and $H = 0.5~\mu$m ] corresponding to relatively small $\Theta_1 < \Theta$ ($\Theta_1 \simeq 140$ and 280, respectively), the $T-V_G$ characteristics are monotonous.
A decrease in $H$ and, hence, an increase in $\Theta_1$  leads 
to steeper $T - V_G$ characteristics.
This is because in shorter GMRs the total  number of the carriers $\propto H$  heated by the inter-GMR current is smaller at the same injected current.

 In contrast, in the case of several  GMR bridges [for example, at $(2N-1) =5$ as shown in Figs.~3(c) and 3(d) with $\Theta_3 > \Theta$], the $T - V_G$ dependences become ambiguous. These dependences are characterized by the threshold voltages ${\overline V}_{3}^{th}$ in the range 50 - 220~mV
 (for the assumed structural parameters). It is instructive that the pertinent values of the threshold
 temperatures (${\overline T}_3^{th} \simeq 33$~meV) only moderately exceed  the lattice temperature $T_0$.
 This implies that in such structures the $J-V_G$ threshold can occur at relatively weak carrier overheating. 
The inclusion of the SC and CC energy relaxation mechanisms results in higher branches of the $T- V_G$
and $J- V_G$ characteristics  leading to the S-shape of these characteristics [see the insets in Figs.~3(c) and 3(d) and in Figs.~4(c) and 4(d)].
The shapes of the  $T-V_G$ and  $J-V_G$ characteristics   shown in  Figs.~3 and 4 are similar . 
The upper branches of these characteristics are characterized by their threshold voltages ${\tilde V}_N^{th} < {\overline V}_N^{th}$. The difference  ${\overline V}_N^{th} -{\tilde V}_N^{th}$ increases with increasing $\Theta_N$ and $\Delta$.

As seen from the comparison of the plots corresponding to different   collision frequency
(the solid and dashed lines in Figs.~3 and 4), the characteristics variations are rather small. This is attributed to  a relatively weak role of the  SC and CC mechanisms 
except in the high voltage/high temperature ranges, at least at the  chosen  PGL parameters (their contribution 
depends on $\nu$). 
Thus, the thermal breakdown and the S-type characteristics can  occur at a large number of the GNR bridges and short GMRs.
It is crucial that in the PGLs under consideration with the identical energy barriers for the electrons injected from the n-GMR and holes injected from p-GNRs, the pertinent injected currents are bound by the positive feedback: the electron injection leads to heating of the 2DHS reinforcing the hole injection and vice versa. 
One needs to stress that the similarity  of the electron and hole injection properties
is  crucial for realizing  the positive feedback in question and the the PGL  S-type characteristics.


For the values of parameters corresponding to the S-type
$J - V_G$ characteristics, the PGL can exhibit two stable current states   with the positive differential conductance  if ${\tilde V}_N^{th} < V_G < {\overline V}_N^{th}$ - low and high current states , ${\overline J}$
and ${\tilde J}$, respectively. 
In the PGLs with the parameters corresponding to Fig.~4(c) and 4(d), $({\tilde J} - {\overline J})/M \sim 100~\mu$A.

The switching between these state can be realized by  voltage pulses
 with the opposite polarity and  the amplitude $\Delta V_G > ( {\overline V}_N^{th} - V_G),
(V_G -{\tilde V}_N^{th})$. The switching time is determined by the processes of the carrier heating and cooling, which are characterized by the room temperature effective energy relaxation time $\tau^{\epsilon}$
being in the range of 10 - 50~ps.
 
 In conclusion, we analyzed carrier heating in PGLs and demonstrated that it can induce hot-carrier thermal breakdown, leading to the S-shaped current-voltage characteristics. This effect arises from the positive feedback between thermionic currents—flowing between neighboring GMRs via GNR bridges—and carrier heating. The unique characteristics of PGLs can enable fast, voltage-controlled current switches operating at room temperature. Furthermore, the features   of the PGL  characteristics make these structures promising for applications such as  radiation detectors.

 The authors are grateful to Prof. T. Otsuji for longtime productive collaboration. 
The work  was supported by  
JST ALCA-NEXT (Grant No.24021835),
NEDO (Grant No.20020912),
Murata Foundation (Grant No.AN24322)
Iketani Foundation (Grant No.0361181-A),
ROHM Co. Ltd, and SCAT Foundation, Japan, the work at RPI
was funded by AFOSR (Contract No. FA9550-19-1-0355).
USA.

\section*{Author Declarations}
The authors have no conflicts to declare.
\section*{Data availability}
The data that support the findings of this study are available
within the article.

\end{document}